\documentclass{article}

\usepackage{arxiv}
\usepackage[utf8]{inputenc} % allow utf-8 input
\usepackage[T1]{fontenc}    % use 8-bit T1 fonts
\usepackage{hyperref}       % hyperlinks
\usepackage{url}            % simple URL typesetting
\usepackage{booktabs}       % professional-quality tables
\usepackage{amsfonts}       % blackboard math symbols
\usepackage{nicefrac}       % compact symbols for 1/2, etc.
\usepackage{microtype}      % microtypography
\usepackage{lipsum}

\usepackage[]{geometry}
\usepackage{mathptmx}
\usepackage{enumerate,cite}
\usepackage[justification=raggedright,singlelinecheck=false]{caption}
\usepackage{graphicx}
\usepackage{amsmath,epstopdf}
\usepackage[subrefformat=parens,labelfont=normalsize]{subcaption}
\usepackage{color,float}
\usepackage{multirow}
\usepackage{blkarray, ulem}

\newcommand{\rn}[1]{{\color{black}#1}}

\usepackage{listings}
\lstdefinestyle{CStyle}{
    basicstyle=\normalsize,
    breakatwhitespace=false,         
    breaklines=false,                 
    captionpos=b,                    
    keepspaces=true,                 
    numbersep=5pt, 
    numbers=left,                  
    showspaces=false,                
    showstringspaces=false,
    showtabs=false,                  
    tabsize=2,
    language=C
}
\makeatletter
\newcommand\newtag[2]{#1\def\@currentlabel{#1}\label{#2}}
\makeatother
\title{BOLD: An Ontology-based Log Debugger for C Programs}

\author{
Dileep Kumar P \\
IIT Madaras \\
Chennai \\
\texttt{dileepkumar\_p@outlook.com} \\
   \And
Rupesh Nasre \\
Department of Computer Science\\
IIT Madaras \\
Chennai \\
\texttt{rupesh@iitm.ac.in}\\
\And
Sreenivasa Kumar P \\
Department of Computer Science\\
IIT Madaras \\
Chennai\\
\texttt{psk@iitm.ac.in}\\
}

\begin{document}

\maketitle
%\IEEEpeerreviewmaketitle
\begin{abstract}
The different activities related to debugging such as program instrumentation, representation of execution trace and analysis of trace are not typically performed in an unified framework. We propose \textit{BOLD}, an Ontology-based Log Debugger to unify and standardize the activities in debugging. The syntactical information of programs can be represented in the from of Resource Description Framework (RDF) triples. Using the BOLD framework, the programs can be automatically instrumented by using declarative specifications over these triples. A salient feature of the framework is to store the execution trace of the program also as RDF triples called \textit{trace triples}. These triples can be queried to implement the common debug operations. The novelty of the framework is to abstract these triples as \textit{spans} for high-level reasoning. A span gives a way of examining the values of a particular variable over certain portion of the program execution. The properties of the spans are defined formally as a Web Ontology Language (OWL) ontology called \textit{Program Debug (PD) Ontology}. Using the span abstraction and PD ontology, end-users can debug a given buggy program in a standard manner. A notable feature of using ontology is that users can accurately debug in some cases of missing information, which can be practically useful. To demonstrate the feasibility of the proposed framework, we have debugged the programs in a standard bug benchmark suite Software-artifact Infrastructure Repository (SIR). Experiments show that the querying time is almost the same as in \texttt{gdb}. The reasoning time depends on the sub-language of OWL. We find that the expressibility offered by OWL-DL language is sufficient for the bugs in SIR programs; but to achieve scalability in reasoning, a restricted OWL-RL language is required.
\end{abstract}


 % \includegraphics[width=\textwidth]{sampleteaser}
  %\caption{Seattle Mariners at Spring Training, 2010.}
  %\Description{Enjoying the baseball game from the third-base  seats. Ichiro Suzuki preparing to bat.}
  %\label{fig:teaser}


\section{Introduction}\label{sec:introduction}

Ontology is a formal knowledge modeling framework that facilitates reasoning and easy integration of information from diversified sources. It has been successfully applied in various domains such as semantic web~\cite{linkeddata}, robotics~\cite{robotics}, genetics~\cite{geneontology}. Today, there is growing interest in ontology-based program analysis~\cite{ecoop,codeontology}. In this paper, we discuss the application of Web Ontology Language (OWL)-based ontology to the debugging problem.

%The software that assists the users in the debugging process is called debugger. 
%There are two types of debuggers: traditional/forward and replay debuggers. The traditional/forward debuggers execute or simulate the execution of the program in the forward direction only. The user can inspect program state of only the current statement being executed. If the user wants to inspect the program state of a previous statement, he has to rerun the program. This problem has been circumvented by replay debuggers. These debuggers allow the users to access the program state of the statements that have already been executed. Depending on the features they provide to the users, these debuggers are further classified into log-based debuggers and reversible debuggers. The log-based debuggers execute the program and store the state information after every instruction in a log. Later in the offline mode, the users can inspect the log to know the state information after the instruction of interest. The reversible debuggers also record the state information after every instruction, but allow the users to inspect the state information in an online mode. If the user is interested in the state of an instruction that has already been executed, these debuggers facilitate the process by replaying the recorded information or reverse executing the program.

The traditional debuggers are reasonably effective in aiding users in identifying bugs. However, those are primarily of \textit{forward} type, and execute or simulate a program only in the forward direction. Thus, users can inspect the state of only the current statement being executed. If users want to inspect the state of a previous statement, they have to rerun the program. To circumvent this issue, two types of debuggers have been proposed~\cite{engblom}: log-based and replay debuggers. These debuggers differ in how the execution trace is stored and reproduced for a query that requires backward-navigation in time. The log-based debuggers~\cite{odb,tralfamadore,pothier,engblomSystem,undodb} execute the program and store the state information in a log prior to the start of the debugging session. For queries that require backward-navigation, these debuggers simply query the log. The replay-based debuggers~\cite{undodb,urdb}, on the other hand, save the results at the instances of the program points where there is non-determinism and create checkpoints. The checkpoints are places where the program state can be safely reconstructed. The backward-navigation queries are answered using checkpoints and saved results.

In this paper, we focus on log-based debugging. It can be easily adapted to complex applications involving multiple threads and machines (e.g., in a distributed setup). It is flexible because they require no human intervention. As the logs are stored, it can be analyzed offline. Kernighan and Pike~\cite{kernighan} advocate that log-based debuggers are more productive than stepping through the code or working with breakpoints. We provide simple, yet practically effective solutions to debugging.

Log-based debugging demands inspection of the execution trace of interest. The trace requires program instrumentation, which can be done at various times~\cite{instrumentationTechniques}: source-code level, intermediate representation level, compilation-time, execution-time instrumentation, and run-time injection. %using binary translation technique
%Various techniques offer trade-offs in instrumentation overheads, platform-independence, compiler-independence, symbol information, etc. Tools may allow a combination of the techniques to improve debugging experience. For instance, \textsf{gdb} can utilize the symbol information obtained from the source code (via \texttt{gcc -g}) to aid a run-time instrumented debugging session.
Since source code continues to be dominant and user-friendly representation of a program, our framework uses source-code instrumentation. 
A variety of techniques exist for source-code instrumentation. The techniques 
%\sout{The} \rn{A} precursor \sout{step} to log-based debugging is to instrument the source code to get the trace information of interest \rn{during program execution}. There are \sout{many different types of instrumentation techniques} \rn{several ways to instrument}~\cite{instrumentationTechniques}: source code instrumentation, intermediate language instrumentation, and compiler assisted techniques such as binary translation, run time instrumentation, and run time injection. In this work we focus \sout{of} \rn{on} source code instrumentation as it is more relevant to debugging. It is implemented in different ways. They 
are classified into preprocessors, tool-specific constructs to define instrumentation specification~\cite{custom1_instrumentation,custom2_instrumentation,custom3_instrumentation}, aspect-oriented programming~\cite{aop_instrumentation,aop_cpp_instrumentation}. A primary drawback of these approaches, except for aspect-oriented instrumentation, is that there is no standard terminology for defining the instrumentation specification. Another limitation is that all these approaches including aspect-oriented instrumentation do not fit into a generalized framework. Thus, the existing approaches are not part of a \textit{unified framework} which solves the instrumentation problem (and other problems discussed below) in debugging in a standardized way.

%the existing debuggers\cite{} store the execution trace information for the programs developed in a particular programming language. This limitation has been addressed by the debuggers\cite{} that store the trace information for multiple languages. But

Several log-based debuggers~\cite{odb,tralfamadore,pothier,engblomSystem} \rn{exist to aid debugging}. \rn{Unfortunately,} these debuggers have two  \rn{shortcomings}. First, the techniques used for storing the trace are \textit{ad hoc}, and limited to the tool-specific infrastructure. There is \textit{no standard, language-agnostic semantic model} for storing the execution trace. This poses a hindrance for establishing communication with other tools that can be used, for instance, to analyze or summarize the trace. A generalized log-based debugger which is independent of the source-code-programming-language and source-code-debugger, but depends only on the trace is very helpful \rn{for software development}. It provides a standard and cost-effective way of debugging applications developed in different programming languages.

The second \rn{shortcoming} is that there are no standard solutions to build abstractions and reason over the trace. This functionality assists the developer to easily diagnose the cause of the bugs. Most of the existing techniques for analyzing the trace are limited to querying or replaying the stored log of the program by simulating the program execution. Some \rn{sophisticated} tools~\cite{expositor,mztake,dalek,ebba,coca} allow defining and executing scripts over the trace, \rn{which are written in high-level languages}. For example, the Expositor~\cite{expositor} and MzTake~\cite{mztake} tools abstract the subsets of the execution trace as lists and event streams respectively. But these approaches lack \rn{a standard} abstraction and hence are limited \rn{only} to the languages supported by these tools. The existing approaches for reasoning are based on \rn{special} libraries available in a programming language, or through an \rn{external} rule based reasoner. For example, the execution trace sub-lists created by the Expositor tool~\cite{expositor} are processed by the list-processing API libraries available in Python. \rn{In effect,} these approaches lack a standard model and \rn{abstraction} useful for reasoning formally. Hence they can produce inconsistent results and fail in case of incomplete information. We propose ontology-based modeling and reasoning to address these limitations.

%he Expositor tool ~\cite{expositor} addresses this issue by abstracting the subsets of the execution trace as lists. But this approach lacks standardization of the terminology for abstraction and hence is limited to the languages supported by gdb. 

Recently, ontologies have been used in declarative program-analysis to standardize the vocabulary and integrate information from different tools. The PATO framework~\cite{ecoop} used ontologies to standardize the syntactic knowledge of C programs. It also showed that ontologies can be used \rn{to} achieve cooperation between different program-analysis-tools using standard vocabulary. The CodeOntology framework~\cite{codeontology} developed ontology for Java programming language. \rn{While these frameworks pioneered the marriage of ontologies and program analysis, they} use ontologies in a \textit{minimal sense}. These frameworks have used ontologies to standardize the vocabulary of a programming language. Further, they use \rn{external} rule-based reasoning systems to carry out the program analysis tasks specified using the standardized vocabulary. 
In effect, these frameworks utilized only the descriptive ability but not the \textit{reasoning power} of ontologies. In this work we \rn{illustrate} both the \rn{standardized} modeling and \rn{the} reasoning benefits of utilizing ontologies for the debugging problem. %We developed three ontologies that differ in the model used to represent the spa

%These frameworks didn't exploit the full potential of ontologies. Some more drawbacks of declarative-program-analysis-systems and debuggers particularly with respect to the reasoning mechanism can be improved by using ontologies.

%The list abstraction is important but not sufficient to address all the debugging needs. For example, . Our intuition is to abstract the trace by means of arbitrary data structures such as trees, lists and graphs and reason over them. [To circumvent this problem, the developers should be allowed to abstract the trace by means of arbitrary data structures. In this work, we explore the benefits of using list, tree and graph data structures to abstract the trace and reason over them.]

%The third drawback is the log

%Thirdly, there are no approaches to control the program instrumentation that fits in the unified framework of the semantic model and analysis of the trace. Program instrumentation is a well-solved problem. However, it will be helpful to unify program instrumentation, semantic model of the trace and reasoning over the trace in one common framework, as these three aspects typically go together.
%We instantiated the framework and developed a system in Java. Specifically, our contributions are the following.

We propose an OWL-based framework called BOLD (Ontology-Based Log Debugger). %to address these drawbacks. 
\rn{We illustrate its utility for debugging C programs.}
In particular, we make the following contributions.
\begin{enumerate}
\item We \rn{introduce BOLD, a framework to} automate program instrumentation based on the C ontology~\cite{ecoop} and the external knowledge expressed as \rn{standard} Resource Description Framework (RDF) triples. This allows programmers to control instrumentation through SPARQL, the W3C recommended query language for RDF.
\item We introduce a semantic model to store the execution traces of C programs as triples. We also facilitate the users of \rn{the} BOLD framework to abstract the trace as \textit{spans} and use the properties of spans for debugging. This reduces both the time and the effort of the programmers in debugging tasks.
\item We propose \textit{Program Debug (PD) Ontology} that is used to standardize the representation of the resources in the trace and the span abstraction. We compare three different versions of the ontology that differ in the model used to represent the span and the sub-language of the ontology.
%\item We propose a semantic model to store the execution trace of C programs. The model is based on Program Debug Ontology (PDO) that is used to standardize the representation of the resources in the trace.
%\item We facilitate the users of BOLD framework to abstract the trace as spans using the vocabulary in PDO and use the properties of span for debugging. This reduces the time and effort of the programmers in debugging tasks.
%We demonstrate it using the list, tree and graph data structures.. We demonstrate it using the list, tree and graph data structures, but it can be extended to other data structures such as stack, heap etc. 
\item  \rn{We illustrate the effectiveness of BOLD using bug benchmark suite Software-artifact Infrastructure Repository (SIR)~\cite{sir}}. Our experiments \rn{reveal} that BOLD is able to diagnose the causes of the bugs using the PD Ontology and OWL reasoner Pellet~\cite{pellet}. Compared to \texttt{gdb}, \rn{BOLD also improves the time taken to query the trace information}.
%\item To..experiment
%The instrumentation is based on \textit{C ontology}~\cite{ecoop} developed for C programming language.

\end{enumerate}
The rest of the paper is organized as follows. Section~\ref{sec:prelims} provides a brief background on RDF, OWL and the associated definitions, \rn{along with} an overview of the conversion of C program into RDF triples using \rn{the} PATO framework~\cite{ecoop}. Section~\ref{sec:bold1} \rn{describes} the program instrumentation and trace model aspects of \rn{the} BOLD framework. Section~\ref{sec:pdo} describes the Program Debug ontology. Section~\ref{sec:obr} \rn{describes the} ontology-based-reasoning \rn{and compares it with} rule-based-reasoning.  Section~\ref{sec:experiments} compares the performance of BOLD to \texttt{gdb}. It also compares the scalability of  OWL reasoners with different ontology models. Section~\ref{sec:relatedWork} compares and contrasts with the existing relevant work. Section~\ref{sec:conclusion} concludes the paper with interesting directions to future work.

%Section~\ref{sec:bold2_debugOperations} explains the different debugging operations supported in \rn{the} BOLD framework.
\section{Background}
\label{sec:prelims}
In computer science, an ontology provides an explicit specification of the shared conceptualization of a domain. It is a description and logic language used to standardize the terminology and knowledge in a domain. The semantic web project\cite{timBernersLee} introduced the languages for realizing ontologies. The description abilities were achieved using the languages such as RDF, RDF Schema (RDFS) and the logic abilities were achieved using OWL. The BOLD framework is developed using these languages. We give an overview of these languages in Section~\ref{sec:swt}. The BOLD framework is built upon the PATO framework~\cite{ecoop} that will be reviewed in Section~\ref{sec:programmodel}.

\begin{figure*}[t!]
%\begin{minipage}[c][10.5cm]{\textwidth}
    %\centering
    \begin{subfigure}[b]{0.35\linewidth}
        \begin{lstlisting}[style=CStyle, escapechar=$]
int maxSubArraySum(int a[], int size) { $\label{mss:function}$
  int globalMax = -32767; $\label{line:globalMaxDecl}$
  int localMax = 0; $\label{line:localMaxDecl}$
  for (int i = 0; i < size; i++) { 
    localMax = localMax + a[i]; $\label{line:localMaxInit}$
    if (globalMax > localMax) $\label{line:globalLocalComp}$
        globalMax = localMax; $\label{line:globalAssignLocal}$
    if (localMax < 0) 
        localMax = 0; 
  } $\label{mss:loopClose}$
  return globalMax; $\label{mss:return}$
}
\end{lstlisting}
        \caption{\normalsize{maxSubArraySum.c}} \label{fig:cfile}
    \end{subfigure}
    \hfill
    \begin{subfigure}[b]{0.5\linewidth}
        \centering
        \[
         \begin{array}{lll}
\textit{file:ln1\_ln12} &\textit{rdf:type} &\textit{c:FunctionDecl}. \\
\textit{file:ln1\_ln12} &\textit{c:hasDefinition} &\textit{file:ln2\_ln12}. \\
\textit{file:ln1\_ln12} &\textit{c:hasReturnType\quad} &\textit{c:int\_type}. \\ 
\textit{file:ln1\_ln12} &\textit{c:hasType} &\textit{c:Function\_type }. \\
\textit{file:ln1\_ln12} &\textit{c:hasName} &'maxSubArraySum'. \\
	% file:3\_1\_3\_31 &rdf:type &\textit{c:ParameterList}. \\
	 \textit{file:ln2} & \textit{rdf:type} &\textit{c:VariableDecl}. \\ 
	 	 \textit{file:ln2} & \textit{c:hasVariable} &\textit{file:ln2Var}. \\ 
	 \textit{file:ln2Var} &\textit{c:hasDataType} &\textit{c:int}. \\ 
	 %file:ln3Var &\textit{c:hasType} &\textit{c:pointer\_type}. \\ 
	 \textit{file:ln2Var} &\textit{c:hasName} &'globalMax'. \\ 
%	\textit{file:ln5} &\textit{rdf:type} &\textit{c:ExpressionStatement} .\\
%\textit{file:ln10} &\textit{rdf:type} &\textit{c:ExpressionStatement} .\\
%\textit{file:ln12} &\textit{rdf:type} &\textit{c:ExpressionStatement} .\\
%\textit{file:ln14} &\textit{rdf:type} &\textit{c:ReturnStatement} .\\
	% file:ln4Var &\textit{c:hasBaseType} &c:\_IO\_FILE. \\
	% file:ln4Var &\textit{c:hasName} &"f". \\
	% file:5\_2\_12\_2 &\textit{c:hasTrueBody} &file:5\_16\_12\_2. \\
	 \text{} & \cdots &\text{} \\
	     \end{array}  
         \]
       \caption{\normalsize{Partial set of triples representing maxSubArraySum.c.}} \label{fig:ttlfile}
    \end{subfigure}

 %   \vspace{.2cm}
   
	\caption{Representation of a C program using triples. The URI reference of the file in Figure \subref{fig:cfile} is abbreviated using the prefix \textit{file} in Figure \subref{fig:ttlfile}.}\label{fig:cprog}
%	\end{minipage}
\end{figure*}

\subsection{Languages for Realizing Ontologies}
\label{sec:swt}
\noindent\textbf{RDF:} The RDF language~\cite{rdf,rdfSyntax} treats every domain as a set of resources that form the domain. In RDF, we represent the information about the resources by making statements about them. Every statement describes a property of a resource and its value. They are formally represented as triples of the form (\textit{subject} \textit{predicate} \textit{object}). Here \textit{subject} is a resource in the domain, \textit{predicate} is a property that has the value denoted by \textit{object}. All the three elements of a triple are  uniquely identified using Uniform Resource Identifiers (URIs). Since URIs are generally long, it is a convention to use abbreviations for them. For example, consider line \ref{line:globalMaxDecl} of the C program given in Figure~\ref{fig:cfile}. RDF treats this variable declaration line as a resource. It can be represented using the URI \textit{http://www.cprograms.com/\allowbreak maxSubArraySum/ln\ref{line:globalMaxDecl}}. We use the abbreviation \texttt{file} to denote the URI \textit{http://www.cprograms.com/\allowbreak maxSubArraySum/}. The variable declaration has many properties such as statement type, name of the variable etc. They are represented as the following triples (\textit{file:ln2 rdf:type c:VariableDecl}), (\textit{file:ln2 c:hasName globalMax}). Here \textit{rdf} is the abbreviation of the URI of the standard RDF terminology recommended by W3C. Similarly, \textit{c} is the abbreviation of the URI of the C ontology proposed in PATO~\cite{ecoop} framework. 

SPARQL~\cite{sparql} is the W3C recommended query language for RDF triples. It is a declarative language that uses triple patterns and operators to specify data retrieval requests on triples. We omit the technical details as they aren't required for this paper.

\noindent\textbf{RDFS:} The vocabulary used to describe the resources and properties in every domain can be standardized to provide uniform representation and avoid ambiguities. The users can use the \textit{standard vocabulary} to make RDF triples. The RDFS~\cite{rdfs} provides a way to declare the standard. It treats every domain in terms of classes/concepts, properties and individuals/instances. A class/concept is used to identify a collection of individuals that share some feature. Note that an individual can belong to multiple classes. The assertion that instance \textit{i} belongs to class \textit{C} is usually written as (\textit{i rdf{:}type C}). A property is used to represent a directed binary relationship between classes or individuals. This is usually written as (\textit{i$_1$ P i$_2$}), where the individual \textit{i$_1$} is related to the individual \textit{i$_2$} through the property \textit{P}. For example, the C ontology developed as part of the PATO framework~\cite{ecoop} is a domain ontology for describing the syntax of C programs. The class \textit{FunctionDefinition} in C ontology contains the function definitions in C programs as the instances. Some other classes in the ontology include \textit{VariableDeclaration}, \textit{ForStatement}, \textit{AssignmentStatement} etc. The property \textit{hasDataType} represents the binary relationship between variables and their data types. Some other properties in the ontology include \textit{hasReturnType}, \textit{hasParent}, \textit{hasScope} etc.

The RDFS language is mainly used to standardize the vocabulary in the domain. It provides constructs to assert the sub-class relationship among classes, domain and range of properties etc. The logic capabilities in RDFS are intentionally restricted \cite{owl}. OWL extends RDFS to represent the logical knowledge and allows reasoning in the domain.

\noindent\textbf{OWL:} The OWL~\cite{owlLanguage} is a logic language that provides operators to define concepts, form concept and property hierarchies. The OWL is a generic term; it is actually of collection of many sub-languages. These sub-languages provide trade-off in terms of expressiveness and reasoning complexity. In this work, we use two sub-languages called OWL-DL and OWL-RL. The OWL-DL language is expressive enough to build knowledge bases of practical significance and provides feasible reasoning. The language is based on Description Logics (DLs). The DLs are fragments of first-order logic that assures the reasoning complexity in feasible time. Since the syntax of OWL is rather verbose, we use the equivalent concise syntax of DLs to describe our domain. The DLs provide operators to define concept expressions from primitive concepts. For example, the $\sqcup$ operator provides union of two concepts and the $\neg$ operator provides the negation (or complement) of a concept, the existential operator $\exists$ acts on a property $P$ to provide the set of individuals in the domain that relate to other via $P$. These concept expression can be used to define new concepts called \textit{defined concepts} using the operator $\equiv$. An example of the definition for the \textit{Adult} class is as follows. \textit{Adult $\equiv$ }(\textit{Male $\sqcup$ Female}) $\sqcap$ ($\exists$\textit{hasAge}[\textit{xsd:long} $>17$]). Adult is defined as a person who is male or female and has age greater than 17.

The complexity of reasoning in OWL-DL is exponential~\cite{dlhandbook}, but for most of the practical requirements the time taken by the reasoners is acceptable. In some rare cases, the reasoners need considerably long time. To circumvent this problem, W3C endorsed three sub-languages as OWL profiles~\cite{owlProfiles}. These sub-languages are less expressive than OWL-DL but reasoning is tractable. We use an OWL-profile called OWL-RL. The features in this sub-language are a subset of the features in OWL-DL. Hence we use the same DL syntax to describe OWL-RL.

\noindent\textbf{Semantic Web Rule Language (SWRL):} There are many assertions of practical interest that can't be represented in OWL-DL\cite{owlrules}. But some of them can be represented using rules. DL-safe SWRL rules~\cite{dlSafeSWRL} can be used with OWL-DL to preserve the feasibility of reasoning. A DL-safe SWRL rule is of the form $atom_1 \wedge atom_2 \wedge \cdots atom_k \rightarrow atom_{k+1} \wedge atom_{k+2} \wedge \cdots atom_{n}$. Each atom is of the form $C(u)$ or $P(v,w)$ where $C$ is a class, $P$ is a property, $u$, $v$ and $w$ are variables or individuals. Informally, the meaning of the rule is "if all the atoms in the antecedent are satisfied for a particular instantiation $I$ of the variables then the atoms in the consequent also hold true for same $I$".

\subsection{Semantic Model of C Programs}
\label{sec:programmodel}
We use PATO framework~\cite{ecoop} to construct a semantic model of C programs. In PATO, C programs are parsed using the ROSE compiler and the RDF triples that describe every line in the program are generated. An example C program and a partial set of the corresponding triples are shown in Figure~\ref{fig:cprog}. The program is a buggy-version of the standard maximum-sub-array-sum algorithm that takes an array of integers and determines the contiguous sub-array in which the sum of the elements is maximum. In the paper when we refer to a particular execution instance of the program, we assume the function is called with the parameters ($\{3, {-5}, 2, 1, 1, {-6}\}$, 6). We use this program and the execution instance with the above input as the running example. In the triples of Figure~\ref{fig:ttlfile}, the subject denotes a construct of the C program in Figure~\ref{fig:cfile}. The prefix \textit{file} is the URI of the C file containing the construct. The meaning of the first triple (\textit{file:ln1\_ln12 rdf:type c:FunctionDecl}) is that the C construct beginning in line~\ref{mss:function} and ending in line~\ref{mss:return} is of type function declaration (\textit{c:FunctionDecl}). The triples following the first triple describe some of the conceptual details of the function declaration statement. For example, the third triple asserts that the return type of the function \textit{file:ln1\_ln12} is of type \texttt{int}. Similarly, some conceptual details of the variable declaration (\textit{int globalMax}) denoted by the URI \textit{file:ln2} are also included in the triple set.%The rest of the triples assert the concept / class of each program statement.

\section{BOLD Framework}
\label{sec:bold1}
The BOLD framework is designed to address the problems of log-based debugging using ontology. It provides solution to the common subtasks of log-based debugging, namely program instrumentation, storage model for the execution-trace, reasoning about and querying over the trace in a unified approach. It is a \textit{unified framework} because all the elements involved in the problem such as program, execution-trace are represented using the same semantic model RDF. Also, the querying and reasoning over RDF are carried out by the standard languages recommended by W3C. In this section, we \rn{describe various} features of the framework. We omit the implementation and user interface details for want of space. %instrumentation and semantic model of the trace. We discuss about debug operations like querying and reasoning in Section~\ref{sec:debugOperations}.

In general, source-code instrumentation \rn{involves} two steps: \rn{(i) identifying source-code statements to be instrumented. The specification for describing these statements is called \textit{instrumentation specification}. (ii) inserting \textit{log-statements} at the statements identified. We describe the instrumentation in Section~\ref{sec:inst_specification}.
%We describe the instrumentation specification in Section~\ref{sec:inst_specification} and program instrumentation in Section~\ref{sec:pi}.
Later, in Section~\ref{sec:tracemodel}, we present the semantic model of the program execution-trace, followed by the description of various debugging operations provided by BOLD in Section~\ref{sec:bold2_debugOperations}.}
%The first step is to identify the statements in the source-code that are to be instrumented.  The specification for describing these statements is called \textit{instrumentation specification}. The second step is to insert \textit{print-statements} at the statements identified in the first step. In this subsection we describe the instrumentation specification, and \rn{defer the explanation of the second step} to the next subsection.

\subsection{Instrumentation Specification}
\label{sec:inst_specification}
%These print-statements capture the trace that can be later analyzed.

%The program instrumentation in BOLD framework is based on Resource Description Framework (RDF) of the ontologies. The services of SPARQL, query language for RDF graphs is availed for achieving the instrumentation objectives. The advantage is that there is a clear delineation between the program to be instrumented and the instrumentation specification. This brings modularity to the process. Unlike the usage of macros recommended by \textit{gcc}, the source-code of the program can be free from the macros that control the instrumentation.

%a \rn{specific} set of
\rn{In general,} it is not feasible to instrument \rn{every statement of} the whole program, as the instrumented program \rn{may} run many times slower than the original ones. So every application that requires instrumentation typically involves identifying the statements to be instrumented. 
\rn{For instance, one may want to instrument on every definition of a variable of interest, and not instrument other statements.}
The requirements of the specification depend on the application \rn{(or client)}. \rn{Based on the source of information feeding into these requirements,} we classify them into two categories. The first \rn{natural} category requires information from \rn{\textit{within}} the source-code. Examples of \rn{such} specifications are identifying the statements where a particular variable is modified, the statements that contain a function call, etc. The second category requires information \textit{outside} the source-code but is relevant for the instrumentation. Examples of \rn{such} external-specifications are identifying the set of functions that belong to a library or the set of functions that "conceptually" belong together
(e.g., all the \rn{variables} that are \rn{vulnerable to} buffer overflow, or \rn{all the} functions that establish a safe TCP connection).

\rn{Existing systems such as \texttt{gcc} advocate changing the source-code of the program (or a header inclusion) which enable specific macros to be available during compilation~\cite{gccMacros}. These macros can then be used in the instrumentation. While this approach works, it is far from the ideal of good software engineering practices. We address this as below.}
In the proposed framework, all the information required for instrumentation specification can be described in the form of triples \rn{(e.g., see Figure~\ref{fig:ttlfile})}. We utilize the source-code-triples generated by PATO~\cite{ecoop} to implement the source-code-specifications. The external information can also be described as RDF triples. For example, all the functions in file library can be asserted as the instances of \textit{FileLibraryFunctions} class. Such identifiers of classes, when used in the specifications, makes them concise and maintainable. The instrumentation specifications can be implemented by running a query to select the statements and the variables / expressions of interest after those statements. We avail the services of SPARQL for information retrieval. The advantage of using SPARQL is that there is a clear delineation between the program to be instrumented and the instrumentation specification\rn{, making the process modular}. \rn{This way,} %Unlike the usage of macros recommended by \texttt{gcc}, 
the source-code of the program can be free from the macros that control the instrumentation.

The BOLD system processes the instrumentation specification given as a SPARQL query. It adds statements to \rn{log} the values of the variables / expressions at appropriate locations. The statements used for logging generate trace information in the form of triples. These triples serve as the semantic model of the trace that we discuss in the following section. 

%The motivation behind the instrumentation specification is as follows. A program instrumentation task typically involves identifying a set of statements and storing the values of the variables/expressions of interest after the statements during the execution. It can be seen that all the information required for specification is available in the  triples of the program. The information can be retrieved by running a query to select the statements and the variable/expressions of interest after those statements. We can avail the services of SPARQL, the W3C recommended query language for RDF graphs for information retrieval. One advantage of using SPARQL is that the variables/expressions to be instrumented can be controlled more effectively in a standard format. For example, the user can filter the subexpressions in a statement or the lvalues of a specific class of statements.

%The second step of the instrumentation process (insert-print-statements)

\subsection{Semantic Model of the Execution Trace}
\label{sec:tracemodel}
The trace information is captured after the execution of the statements as dictated by the instrumentation specification. The information includes the variable values and pointer addresses after the statements of interest. Note that this information can be captured after the same statement $S$ multiple times during the execution. It is important to distinguish between the multiple visits of the statement $S$. To fix this issue, we associate \rn{a unique} \textit{timestamp} with the information captured after each statement. The timestamps are implemented as natural numbers and follow a linear order with the relation $<$. If the trace information captured after two statements $S_1$ and $S_2$ has timestamps $t_1$ and $t_2$ respectively and $t_1 < t_2$, then $S_1$ \rn{must have been} executed before $S_2$ \rn{(and vice versa)}.

\newcommand{\tstamp}[1]{{\textbf{\color{black}{#1}}}}
\newcommand{\globalMaxSpanHighlight}[1]{{\textbf{\color{black}{#1}}}}
\newcommand{\spanHighlight}[1]{{\textbf{\color{blue}{#1}}}}

We define the trace information captured after each statement as a model called \textit{execution-trace model}. The model is same for scalar variables, expressions, array member accesses and pointers to scalar data types. Even though the model is same for all the above mentioned types of variables, we refer to the model as simply the model of scalar variables for simplicity of writing. Intuitively, the model for scalar variables is described by the tuple $\langle v_{id}, s_{id},  val, \tstamp{timestamp}\rangle$. The element $v_{id}$ is a variable that has the value $val$ after the statement $s_{id}$ during the execution. The element \textit{\tstamp{timestamp}} identifies the position of visit of $s_{id}$ in the overall sequence of visits of all statements during the execution. The model slightly changes for the member variables of \texttt{struct} data type in C. For these variables, the model is described by the tuple $\langle v_{id}, decl_{id}, s_{id},  val, \tstamp{timestamp} \rangle$. The element $v_{id}$ is a member of the \texttt{struct} data type. The element $decl_{id}$ identifies the variable of the \texttt{struct} data type used to access the member $v_{id}$. The remaining elements are \rn{the} same as \rn{those} of the model described for scalars. The trace information captured after all the statements can be conceived of as one big list. We call this list as \textit{execution-trace list}.

Recall that we use the input array $\{3, {-5}, 2, 1, 1, {-6}\}$ to capture the execution trace for the C program in Figure~\ref{fig:cprog}. The partial set of \textit{execution-trace model} is shown in Figure~\ref{fig:semTraceModel}. The tuples describe the trace information of the variable \textit{localMax} (identified by the URI (\textit{file:ln3Var}))  after line~\ref{line:localMaxInit} in the program. The tuples are captured in different loop iterations. This is indicated by the tuple's last element \textit{\tstamp{timestamp}}, which is in the increasing order. The third element denotes the value of the \textit{localMax} variable in different iterations.
%\dileeptodo{Would the information be more accessible if we use a couple of colors to designate certain variables across figures?}
%zzz: run the program and adjust the timestamps
\begin{figure}
%\begin{minipage}[c][10.5cm]{\textwidth}
   \centering
   
        \[
         \begin{array}{l}
(\textit{file:ln3Var, file:ln5, 3, \tstamp{3}}) \\
(\textit{file:ln3Var, file:ln5, -2, \tstamp{6}})\\
(\textit{file:ln3Var, file:ln5, 2, \tstamp{9}}) \\
(\textit{file:ln3Var, file:ln5, 3, \tstamp{14}})\\
(\textit{file:ln3Var, file:ln5, 4, \tstamp{17}})\\
(\textit{file:ln3Var, file:ln5, -2, \tstamp{20}})\\

	     \end{array}  
         \]
         %\captionsetup{aboveskip=2pt,belowskip=2pt}
	\caption{Partial set of the \textit{execution-trace model} captured after line~\ref{line:localMaxInit} of C program in Figure ~\ref{fig:cprog}. The identifier \textit{file:ln3Var} denotes the variable \textit{localMax} in the program}\label{fig:semTraceModel}
%	\end{minipage}
\end{figure}

%\dileeptodo{This paragraph seems redundant to me.}
%The \textit{execution-trace model} captured after each statement can be sorted based on timestamp and the resultant can be conceived of as one big list. We call this list as \textit{execution-trace list}. This list can be conceived as an instance of list data structure  encountered in different programming languages- the java.util.List class in Java, the System.Collection.Generic.List in C\#, the Data.List module in Haskell.

% The timestamp of the execution is denoted by \textit{timestamp}.

%The semantic model used to store the trace is based on RDF.  The vocabulary used in the triples is based on Program Analysis ontology that we designed.  

%\subsection{Motivation for abstraction-debug-operation}
%\label{sec:motive_abstraction}
%The BOLD framework supports three types of debug operations on the trace: querying over the trace, high-level abstractions over the trace and reasoning over the abstractions. These operations are elaborated in the next section. The novel contribution of this work is the introduction of abstraction operation in debugging. In this section, we discuss motivation and advantages of including this operation. We discuss it in the broader context of program analysis, but the advantages hold for debugging too.

\subsection{Debugging Operations in BOLD}
\label{sec:bold2_debugOperations}
%Users perfrom two operatiosn:  querying and reasoning. user issues sparql, get the result. To perfrom reasoning, user issues sparql, get relevant data, convert it to a suitable data model that is expressed in the form of triples, and reason using the common data model operations.

The BOLD framework supports three types of debug operations on the trace: querying the trace, creating high-level abstractions over the trace, and reasoning over the created abstractions. To facilitate debugging, the framework allows users to have an interactive debugging session with the system similar to \textsf{gdb} session. During the interaction, the users can issue \rn{various framework} commands. These commands are used to invoke the functions which perform the debug operations. The syntax is similar to Datalog atoms.

\subsubsection{Querying the Execution Trace}
\label{sec:querying}
%\dileeptodo{A reader won't be interested in knowing about various types of queries, unless we provide a concrete example. Let's add a single example. If it can be the same as in the previous section, that would be wonderful. Let's use it as a running example to explain almost every aspect of BOLD.}

%Querying the information available in execution-trace is a standard feature offered by the debuggers. We classify the queries available in BOLD framework into three classes: \textit{standard} queries, \textit{aggregate} queries, and \textit{integrated} queries. We describe these query types below. All the queries are posed to the system through direct-query-commands. Recall that these commands refer to SPARQL queries. The standard and the aggregate queries run only on the execution-trace triples. The integrated queries run on the cumulative triple store formed from the integration of source-code triples, execution-trace triples, and external triples (explained in Section~\ref{sec:inst_specification}). In the rest of the section, we explain about these queries.

Querying the information available in the execution-trace is a standard feature offered by the debuggers. The queries are posed to the system as framework commands are essentially SPARQL queries. The queries \rn{run on the execution-trace triples and} are used to perform the commonly used debug operations such as \texttt{break} (to go to a desired statement), \texttt{inspect} (to know the information of the variables at a particular statement), and \texttt{step} (to step-through instructions). Note that the existing debuggers cannot express source-code or external-knowledge (relevant knowledge outside the program) related requirements in a standard way. For instance they can't express the computation of the execution path in a function, the size of an array after the class of functions that can cause buffer-overflow, or the value of file pointer after file library functions. The reason is that the source-code knowledge, and external knowledge are typically not available together in a standardized way to the debuggers. We address this issue in BOLD framework through a class of queries called \textbf{integrated queries}. They are called integrated because they run on the cumulative triples formed from the integration of three different triple stores, namely, source-code triples, execution-trace triples, and external triples. The integration of knowledge from multiple sources is an easy task in ontology because of the use of URIs to represent the resources.

\subsubsection{Abstractions of the Execution Trace}
\label{sec:abstraction}
%During implementation, the \textit{execution-trace model} captured after each statement is sorted based on timestamp and stored in a list called \textit{execution-trace list}. %

 %The \textit{execution-trace list} is abstracted by means of different data models such as lists, trees and graphs to ease the debugging tasks. The abstractions  extract useful information from the execution-trace list and represent it in the form of a data model. By using ontologies for representing the data models, the abstraction is done in a debugging-tool-independent and programming language-independent way.
%Recall from Section~\ref{sec:obr} that the abstraction phase is the preprocessing step to use ontology-based-reasoners. 
The abstraction phase is used to create meaningful abstractions of the trace that provide effective reasoning. In BOLD framework, it is defined as the process of constructing new sequences (called \textit{spans}) from the execution-trace list. Each \textit{span} contains the information related to a variable or a simple expression and is expected to satisfy a property. We explain about spans in this section and their properties in Section~\ref{sec:reasoningImplementation} %\rn{We explain about spans in more detail in this section and the properties of spans in Section~\ref{sec:reasoningImplementation}}. %(discussed in Section~\ref{sec:reasoningImplementation}). %\dileeptodo{Do you mean trace-list? This list model is not coming out clearly. we need to explain it better.}
 
Each span contains the values of a variable or an expression retrieved from the execution-trace list at discrete time stamps. The span is formally described as $\langle name, var, seq\_of\_cells \rangle$. Here $name$ is the name of the span and $var$ is the variable or the arithmetic-expression \rn{for} which the span is constructed. $seq\_of\_cells$ is \rn{an ordered list} of cells, where each cell is described as $\langle val,\text{ } timestamp \rangle$. The $val$ element is the value of the variable $var$ at the timestamp identified by the element $timestamp$. To facilitate reasoning in the standard format, \rn{BOLD stores} the spans in the form of RDF triples. 
 
 %As an example, for the program in Figure~\ref{fig:cfile}
 
In the running example, to record the variable values at the end of \rn{each iteration of} the \texttt{for}-loop, we construct a span \rn{each} for variables \texttt{globalMax}, \texttt{localMax}, and \texttt{i}. One can see from the (intended) semantics of the program, that the \rn{values stored in the respective} span would be non-decreasing for \texttt{globalMax}, non-negative for \texttt{localMax}, and strictly increasing for the variable \texttt{i}.
 
To construct a span, the users pose framework commands related to abstraction to the BOLD system. The system executes these commands to retrieve the timestamps called \textit{span-timestamps}. Timestamp is a system-internal element and the users don't have access to it. Therefore, one of the interesting challenges in BOLD is to allow users specify the timestamps. Recall from Section~\ref{sec:tracemodel} that timestamp is associated with the execution of \textit{each instance} of the statement \rn{during execution}. One solution that BOLD system implements is to let the users specify the statement identifiers; so that the system can retrieve timestamps from them.

The formal specification of the abstraction commands is described as $\langle var_{span}, st_{span}, \phi \rangle$.
Here $var_{span}$ is the variable for which the span is constructed. The element $st_{span}$ is the identifier of the statement after which the variable $var_{span}$ is accessed. Since the statement $st_{span}$ may be executed many times, it is necessary to filter and identify the required instances. We \rn{resolve this issue by introducing} the concept of \textit{filters}. The last element in the specification $\phi$, called filter, is used to identify the instances.
%Here the values of the variable $var_{span}$ after the statement $st_{span}$ form the span. The element $\phi$ is the set of timestamps, called \textit{span-timestamps}, 

%These commands demand the ability to specify timestamps -- as the user may be interested in a particular instance of a variable. 
In the running example, we construct a span on \texttt{globalMax} variable that contains the values of the variable \textit{at the end} of the \texttt{for}-loop in different iterations. We refer to this span as \texttt{\globalMaxSpanHighlight{globalMaxSpan}}. Similarly, we construct a span on \texttt{globalMax} variable at the end of the \texttt{for}-loop in the iterations where  \spanHighlight{the value of \texttt{localMax} after line~\ref{line:localMaxInit} is positive}. We refer to this span as \texttt{\spanHighlight{globalMaxFilteredSpan}} and the condition highlighted in blue as \texttt{\spanHighlight{localMaxPositive}} condition. Next we explain the details of filters.

%This span contains the values of the variable \texttt{globalMax} in different iterations of the loop.
%globalMax\allowbreak Filtered\allowbreak Span\allowbreak Condition
 %The interesting part here is to provide appropriate commands to specify the timestamps. (say, variable \texttt{globalMax} in the third iteration or all instances beyond the third iteration)

%We call the timestamps at which a span is constructed as \textit{span-timestamps}. Timestamp is a system-internal element and the user is not expected to have knowledge about it. Therefore, one of the interesting challenges in BOLD is to allow users specify the timestamps. Recall from Section~\ref{sec:tracemodel} that timestamp is associated with the execution of \textit{each instance} of the statement \rn{during execution}. One solution that the BOLD system implements is to allow users to specify the statement identifiers; so that the system can compute timestamps from them. However, since a statement may be executed many times, it is still necessary to filter and identify the required instances of a statement. % and their associated span-timestamps.
 
  %in terms of statement identifiers and their instances
  %Next we explain about the types of filters.  The filters defined as intervals and sets are called filter-intervals and filter-sets respectively.
%We \rn{resolve this issue by introducing} the concept of \textit{filters} to identify the particular instances of statements. 
A filter is formally defined as an interval or a set of timestamps relative to which the span-timestamps are computed. The filters are of two types: interval-based filters and set-based filters. An interval-based filter is specified using two timestamps: lower and upper bounds of the interval. Both the bounds are optional. If they are not specified the BOLD system considers the lower and upper bounds as $0$ and $\infty$ respectively. It considers only the timestamps that are with the bounds of the interval to construct the spans. Intuitively, a filter-interval specifies a sub-list of the execution-trace list over which the spans are constructed.

%If the abstraction commands contain a filter-interval,

The motivating example for set-based filters is as follows. %Recall that \texttt{globalMaxFilteredSpan} is constructed on \texttt{globalMax} variable at the end of the \texttt{for}-loop in the iterations where \textit{the value of \texttt{localMax} after line~\ref{line:localMaxInit} is positive}.
In the \texttt{\spanHighlight{globalMaxFilteredSpan}} specification, the \texttt{\spanHighlight{localMaxPositive}} condition is true after many instances of the statement in line~\ref{line:localMaxInit} because the statement is in a loop. In the iterations where the value of \texttt{localMax} is positive, we are interested in the value of \texttt{globalMax} at the end of the loop. Note that the timestamp at the end of the loop is different from the timestamp after line~\ref{line:localMaxInit}. To construct the \texttt{globalMaxFilteredSpan}-like spans, we introduce the concept of set-based filters. A set-based filter is described as a set of timestamps of the form $\phi = \{t_1, t_2, \cdots, t_n\}$. The BOLD system interprets each of the successive elements of the $\phi$ set as a pair. So it forms a set of pairs of timestamps $PAIRS = \{(t_1, t_2), (t_2, t_3), \cdots, (t_{n-1}, t_n), (t_n, \infty)\}$ for each filter-set. Each pair in the $PAIRS$ set provides one span-timestamp (the first such span-timestamp if there are many). The span-timestamp lies within the bounds of the pair. In the \texttt{\spanHighlight{globalMaxFilteredSpan}} specification, the timestamps related to line~\ref{line:localMaxInit} where the \texttt{\spanHighlight{localMax\allowbreak Positive}} condition is true form the $\phi$ set. The timestamps at the end of the loop in those iterations where the condition is true form the span-timestamps.

%For example, the \texttt{globalMaxFilteredSpan} described above is defined using filter-set. The filter-set is defined using the state of the variable \texttt{localMax}. The timestamps where the value of \texttt{localMax} after line~\ref{line:localMaxInit} is positive form the $FTS$ set. The values of \texttt{globalMax} after line~\ref{line:globalAssignLocal} in each pair of the $FPTS$ set form the \texttt{globalMaxFilteredSpan}.

%If the abstraction commands contain a filter-interval, then 

%The filters are specified by the users as part of the abstraction commands. So it is reasonable to expect that they are
The lower, upper bounds of interval-based filters and the elements in filter set are timestamps. As already mentioned, the users don't have access to the timestamps. But they have to specify them as part of the abstraction commands. We tackle this issue by specifying them in terms of the states of the variables (called filter-variables). The state of a variable is formally described as a 4-component tuple $\langle s_{id}, v_{id}, rel, val\rangle$. The meaning of the tuple is that the relation element $rel$ describes the equality / inequality relationship between the filter-variable $v_{id}$ and the value $val$ \textit{after} the statement $s_{id}$. Note that the filter-variables can be different from the variable for which the span is constructed. For example, in the \texttt{\spanHighlight{globalMaxFilteredSpan}} specification, \texttt{localMax} is the filter-variable and \texttt{globalMax} is the variable for which the span in constructed. The statement instances in the filter can be same as that of the statement instances at which the span is constructed or any nearby statements' instances. In \texttt{\spanHighlight{globalMaxFilteredSpan}} example, the statement identified by \textit{file:ln5} is used to define the filter and \textit{file:ln8-ln9} (RDF identifier for the if-conditional in lines 8--9 of Figure~\ref{fig:cfile}) is used to construct the span. Using the above explanation, the \texttt{\spanHighlight{globalMax\allowbreak Filtered\allowbreak Span}} is specified by the tuple $\langle$\textit{file:ln2Var}, \textit{file:ln8-ln9}, $\langle$\textit{file:ln5}, \textit{file:ln3Var}, $>, 0 \rangle \rangle$.

\section{Program Debug Ontology}
\label{sec:pdo}

\begin{figure*}[t!]
\captionsetup[subfigure]{singlelinecheck = false, justification=raggedright}
%    \captionsetup[subfigure]{position=b}
%\begin{minipage}[c][10.5cm]{\textwidth}
   \begin{subfigure}[t]{0.3\textwidth}
   \centering\[
         \begin{array}{lll}
\textit{file:ln3Var} & \textit{pd:hasState} & \textit{pd:st1} \\
\textit{pd:st1} & \textit{pd:afterStatement} & \textit{file:ln5}\\
\textit{pd:st1} &\textit{pd:timeStamp} &\textit{3}\\
\textit{pd:st1} &\textit{pd:value} &\textit{3}\\
%\text{} & \cdots &\text{}\\
%\\~\\

	     \end{array}  
         \]
     \captionsetup{skip=26pt}
    \caption{\normalsize{The tuple (\textit{file:ln3Var, file:ln5, 3, 3}}) of the execution trace from Figure~\ref{fig:semTraceModel}}\label{fig:trace_triples}
	\end{subfigure}\hfill
	\begin{subfigure}[t]{0.3\textwidth}
        \centering\[
        \begin{blockarray}{lll}
         \textit{pd:l1} & \textit{rdf:type} & \textit{pd:Span} \\
          \begin{block}{\{>{\medspace}lll}
         \textit{pd:l1} & \textit{rdf:type} & \textit{rdf:List} \\
         \textit{pd:l1} & \textit{rdf:first} & \textit{-32767} \\
         \textit{pd:l1} & \textit{pd:timeStamp} & \textit{5} \\
         \textit{pd:l1} & \textit{rdf:rest} & \textit{pd:l2} \\ 
        \end{block}
	 \text{} & \cdots &\text{} 
      \end{blockarray}
      \]
      \captionsetup{skip=-5pt}
       \caption{\normalsize{The first cell of the \texttt{\globalMaxSpanHighlight{globalMaxSpan}} abstraction with list model}}\label{fig:triples_listmodel}
    \end{subfigure}\hfill
	\begin{subfigure}[t]{0.3\textwidth}
        \centering\[
        \begin{blockarray}{lll}
         \textit{pd:l1} & \textit{rdf:type} & \textit{pd:Span} \\
          \begin{block}{\{>{\medspace}lll}
         \textit{pd:l1} & \textit{pd:hasSpanCell} & \textit{pd:c1} \\
         \textit{pd:c1} & \textit{pd:hasValue} & \textit{-32767} \\
         \textit{pd:c1} & \textit{pd:timeStamp} & \textit{5} \\
         \textit{pd:c1} & \textit{pd:index} & \textit{0} \\
        \end{block}
	 \text{} & \cdots &\text{} 
      \end{blockarray}
      \]
      \captionsetup{skip=-5pt}
       \caption{\normalsize{The first cell of the \texttt{\globalMaxSpanHighlight{globalMaxSpan}} abstraction with set model}}\label{fig:triples_setmodel}
    \end{subfigure}
    %\captionsetup{skip=-5pt}
       \caption{\normalsize{Partial set of triples representing the execution trace and the abstractions \rn{for our example}}}\label{fig:triples_complete}
\end{figure*}

Our Program Debug Ontology provides the vocabulary and axioms to standardize the information generated in different features of the BOLD framework. In this section, we explain the important terms of the ontology. The execution trace and the span abstraction defined above are represented and stored in the form of triples. The ontology also provides axioms and rules that formalize the reasoning process in BOLD, \rn{which we explain in the next section.}

Recall that the execution-trace model for variables is described by the tuple $\langle v_{id}, s_{id},  val, timestamp\rangle$ that maps the variables to different states. This tuple is stored as a set of triples. The property \textit{hasState} maps a variable to a state. The state is elaborated using three properties. The value of the first property, \textit{afterStatement}, provides the statement id $s_{id}$ after which the state is captured. The second property, \textit{hasValue}, provides the value of the variable after the statement $s_{id}$. The value of the final property, \textit{timestamp}, provides the time-stamp value. \rn{For our running example,} the triples for the first tuple (\textit{file:ln3Var, file:ln5, 3, 3}) in Figure~\ref{fig:semTraceModel} are presented in Figure~\ref{fig:trace_triples}. The individual \textit{pd:st1} denotes the state of the variable after executing the statement \textit{file:ln5} at time-stamp 3. 

%in the execution sequence

The span abstraction \rn{is also} represented in the form of triples. We implemented two models to realize the triples corresponding to spans. They are the \textit{list model} and the \textit{set model}. The elements in the list model are ordered based on time-stamp, whereas the elements in the set model are unordered. In principle, they both are equivalent because the information present in both the models is the same. The difference arises in the time taken to realize different properties of the span. We will empirically show it in Section~\ref{sec:experiments}. 

%\dileeptodo{list would also have duplicates?}

We used the standard \textit{rdf:List} to implement the list-model of spans. The \textit{rdf:List} is a recursive list data structure that is in principle similar to lists in Prolog. It contains a sequence of cells connected by the property \textit{rdf:rest}. The \rn{RDF} standard represents the contents of each cell using the property \textit{rdf:first}. In BOLD, the content of a cell is the value of a variable at time-stamp $t_0$. Additionally, we used the property \textit{timeStamp} to assert the time-stamp value $t_0$ of a cell. The set-model is similar to the list-model except that the cells are not explicitly connected to each other. Instead, every cell has an additional property called \textit{index} which gets used to maintain the relative position in the span. %\dileeptodo{don't we store actual timestamp value in the set cell? If we don't, how do we search for a timestamp? If we do, why do we need an index?}

%\dileeptodo{Let's recall the definition, or if it is not used earlier, let's define it here itself.} 
Recall that, in our running example, \texttt{\globalMaxSpanHighlight{globalMaxSpan}} contains the values of the \texttt{globalMax} variable in different iterations of the \texttt{for}-loop. The \texttt{\globalMaxSpanHighlight{globalMaxSpan}} has a constant value -32767 in all the cells (because of a bug that will be identified in the next section). The RDF representation of the first cell in the span in both the list and the set models is presented in Figure~\ref{fig:triples_listmodel} and Figure~\ref{fig:triples_setmodel} respectively. In both the figures, the span is identified by the individual \textit{pd:l1}. The first cell is identified by \textit{pd:l1} in the list model and \textit{pd:c1} in the set model. Note that the identifiers of the span and the first cell are the same in the list model because list model is represented by a sequence of cells that are linked to each other (analogous to singly linked list). The linkage is done through the property \textit{rdf:rest}. In the Figure~\ref{fig:triples_listmodel}, the last triple shows the link between the first and the second cells. When there are no more links, the \textit{rdf:nil} individual is used to terminate the list. There are no such links in the set model. Instead, the index of the cells in represented using the property \textit{pd:index}. In both the figures, the triples given in the curly braces describe the contents of the first cell. Similar set of triples \rn{is} used to represent every other cell. \rn{This makes the representation standard, and querying and reasoning concise.} The concepts of the ontology that aid in debugging are presented in Section~\ref{sec:reasoningImplementation}.

%\dileeptodo{A debug ontology needs more explanation. how different debugging concepts are captured in the ontology etc.}

%It is similar in functionality to lists in Prolog as both of them split the list into head and tail. 
\section{Ontology-based reasoning}
\label{sec:obr}
%We utilize Ontology-based reasoning in BOLD framework. The Ontology-based reasoning is a type of declarative reasoning but is differe

% Even though it is popular for the productivity advantages, it has certain limitations like non-standardization of the vocabulary, no standard way of integrating information from different tools etc. certain limitations have been addressed by the usage of ontologies. 

%The declarative-program-analysis is a paradigm of program analysis based on user-defined vocabulary and logic rules to specify the analysis. Even though it is popular for the productivity advantages, it has certain limitations like non-standardization of the vocabulary, no standard way of integrating information from different tools etc. Recently PATO framework~\cite{ecoop} and CodeOntology framework~\cite{codeontology} used ontologies to address these limitations. These frameworks used ontologies only as a descriptive language and not as a logic language. They used 
Reasoning in the BOLD framework is the process of inferring the properties of spans created in the abstraction phase.
There are two approaches to implement reasoning in the debugging systems. They are ontology-based-reasoning and rule-based-reasoning. Most of the existing declarative-debugging systems \cite{coca,declarative_debugging1,declarative_debugging2} are based on rule-based-reasoning. The novel contribution of this work is the usage of ontology-based reasoning for debugging. %Section~\ref{sec:comparison} compares the strengths and limitations of ontology-based-reasoning against rule-based-reasoning. Section~\ref{sec:reasoningImplementation} discusses the implementation of reasoning in BOLD.

We discuss three advantages of using ontology-based reasoning compared to rule-based reasoning. Foremost, ontologies provide a standard way to define non-primitive concepts in terms of primitive concepts, properties, and individuals using the operators. So by using an ontology-based system, the vocabulary used in debugging can be standardized. Second, the ontology-based systems are designed to always produce consistent inferences. Such systems generate inferences only if the ontology axioms and rules are consistent. In contrast, the rule-based systems can not detect inconsistencies in the rule specifications. Rules can be accidentally written to infer both an assertion and its negation. Third, the ontology-based-reasoning systems can produce accurate results in some cases of incomplete information, which can be practically useful. This is because they adopt \textit{open-world} assumption. This means the facts \rn{that} are unavailable are considered to be \textit{unknown}. We explain this difference with an example.

\begin{figure}[t!]
%\begin{minipage}[c][10.5cm]{\textwidth}
    %\centering
   % \begin{subfigure}[b]{0.35\linewidth}
        \begin{lstlisting}[style=CStyle]
int main() {
	int A[] = {5, 9, 2, 4};
	f(A); // A is sorted in ascending order
	g(A); // The "sorted" order of A is unknown
	h(A); // A is sorted in non-ascending order
}
\end{lstlisting}
%\captionsetup{aboveskip=2pt, belowskip=2pt, skip=5pt}
        \caption{A partial C program to illustrate the advantages of ontology-based-reasoners} \label{fig:cfile_obr}
    \end{figure}

Consider a partial C program presented in Figure~\ref{fig:cfile_obr}. All the functions \textit{f}, \textit{g} and \textit{h} sort the array that is provided as the parameter. The program is executed and the trace contains the values of array \textit{A} after \textit{f} and \textit{h} function calls but does not contain information about \textit{A} \rn{immediately} after \textit{g}. After the function calls \textit{f} and \textit{h}, the values of array \textit{A} are in ascending and non-ascending orders respectively. From the intended semantics of the program, it is known that \textit{A} is sorted after function call \textit{g}. But whether \textit{A} is in ascending or non-ascending order is unknown. Suppose a query is posed to the reasoner asking whether the main function contains consecutive function call statements after which \textit{A} is either in ascending or non-ascending orders. As the information about \textit{A} is unavailable after \textit{g}, the rule-based-reasoners can not find consecutive calls that satisfy the query. Hence, they return \textit{no} as the answer. The ontology-based-reasoners can utilize the fact that every sorted array must be in either ascending or non-ascending order (can be provided through ontology). So \textit{A} must be in one of the two orders after \textit{g}. It then computes the answer based on cases. In the first case where \textit{A} is assumed to be ascending, the reasoner has found consecutive function calls \textit{f} and \textit{g}. In the second case where \textit{A} is assumed to be non-ascending, the reasoner has found consecutive function calls \textit{g} and \textit{h}. Thus, the reasoner returns 
the accurate answer (\textit{yes}).
%\textit{yes} as the answer, \rn{which is accurate.}

%\rn{By non-ascending, you mean randomly ordered? If yes, it is confusing. If you meant descending, that need not be the case with every array.} 

Ontology-based reasoning systems \rn{also suffer from certain drawbacks}. The first drawback is that these systems don't allow new individuals to be created during the reasoning process to maintain computational feasibility. \rn{In practice, however, this is not a deterrent, especially for procedural style programs. This is because in program analyses, we often need to store the intermediate results.} For example, if we are reasoning over tree data structure, we might store the result of a traversal. \rn{Therefore, to adddress this drawback,} we propose a preprocessing step in BOLD \rn{which} creates new individuals \rn{to} be utilized by ontology reasoners. The abstraction phase discussed in Section \ref{sec:abstraction} is the preprocessing step. 

 %This step can be implemented either in rule-based programming or imperative programming. 
%standardization
%consistency in reasoning
%Reasoning about unground individuals whose existence is known
%sound reasoning (prolog has unsound reasoning in case of errors.)

 %W3C has standardized restricted versions of OWL that support scalable reasoning. These versions are called \textit{OWL profiles}.

The second drawback of \rn{ontology-based systems} is that, as already mention\rn{ed} in Section~\ref{sec:prelims}, the OWL-DL reasoners are exponential time algorithms~\cite{dlhandbook}. It means the reasoners are not scalable with increase in the size of data. We illustrate the limits empirically in Section~\ref{sec:experiments}. To overcome this drawback, we expressed the reasoning properties used for debugging in OWL-RL profile (refer to Section~\ref{sec:reasoningImplementation}). As we will show in Section~\ref{sec:experiments}, the ontology expressed in OWL-RL scales better than the one expressed in OWL-DL.

%\vspace{-20pt}
\subsection{Implementation in BOLD}
\label{sec:reasoningImplementation}
Identifying span properties is helpful to determine the cause of the program bugs. \rn{We discuss about two types of properties:} \textit{intra-span} and \textit{inter-span}. The intra-span properties are useful to verify the properties of the values of a variable abstracted into a span. This abstraction is mostly done after a particular statement in different iterations of a loop, or in a sequence of statements in a program, or relative to the statements that hold a particular value of a variable. The intra-span properties perform universal, existential, and comparison checks on the span. An example of universal check is whether all the elements are positive. Similarly, one can check for negative, zero, non-positive, non-negative, and non-zero elements. An example of existential check is whether the span contains a positive element. Similarly, one can check for a negative element, a zero element, duplicate elements and unique elements. The comparison checks verify "sorted"-ness properties of a span such as ascending, descending, non-ascending and non-descending. In our running example, we can verify the properties of \texttt{\globalMaxSpanHighlight{globalMaxSpan}} and \texttt{\spanHighlight{globalMaxFilteredSpan}} constructed on \texttt{globalMax} variable. Recall that \texttt{\globalMaxSpanHighlight{globalMaxSpan}} contains the values of the variable in all iterations of the \texttt{for}-loop and \texttt{\spanHighlight{globalMaxFilteredSpan}} contains the values of the same variable in the iterations where the \texttt{localMax} variable is positive. If the program is correct, the \globalMaxSpanHighlight{globalMaxSpan} must be a non-descending span and \texttt{\spanHighlight{globalMaxFilteredSpan}} must be a span with all the positive elements. In the running example, the first property is satisfied and the second property is violated. It means even though the value of \texttt{localMax} is positive, \texttt{globalMax} is not updated. Hence there must be a bug in lines \ref{line:globalLocalComp}--\ref{line:globalAssignLocal} related to \texttt{globalMax} variable assignment.

%\rn{Examples of the universal checks are:} whether all the elements are positive or negative or zero or non-positive or non-negative or non-zero. \rn{Examples of the existential checks are:} whether the span contains a positive element or negative element or zero element or duplicate elements or \rn{unique} elements.

The inter-span properties are useful to verify the properties related to two different spans. Note that the two spans can belong to different variables / expressions. They are useful to verify loop invariants, assertions described as \texttt{if-else} rules, relations between variables etc. They are most likely to be verified within a block of statements such as a particular loop or a particular instance of a function call. We address this issue using the idea of \textit{comparable spans}. Let the sequence of cells in two spans $S_1$ and $S_2$ be described as $\{\langle val_{11}, t_{11} \rangle, \langle val_{12}, t_{12} \rangle, \cdots, \allowbreak \langle val_{1n}, t_{1n} \rangle\}$ and $\{\langle val_{21}, t_{21} \rangle, \langle val_{22}, t_{22} \rangle, \cdots, \langle val_{2n}, t_{2n} \rangle\}$, where $val_{ij}$ is the value in span $i$ at timestamp $t_{ij}$. Two spans $S_1$ and $S_2$ are \textit{comparable} if $t_{1i} \leq t_{2i} < t_{1(i+1)}$ for $i = 1 \text{ to }n-1$ and $t_{1n} \leq t_{2n}$. Intuitively, in two comparable spans $S_1$ and $S_2$, the cell at index $i$ of $S_2$ is ordered based on the timestamp between the cells at index $i$ and $i+1$ of $S_1$. In BOLD framework, to perform coherent operations, we restricted that all inter-span properties be verified only on \textit{comparable spans}. The inter-span properties are used to verify the relation between the values in two spans at the same index position. The relationships can be any of the standard comparison operators such as $=, !=, <, <=, >, >=$. In the running example, we can define a span on the \texttt{localMax} variable that contains the values of the variable after line ~\ref{line:localMaxInit} in different iterations of the loop. If the program is correct, each value in this span is less than or equal to the same-index value in \texttt{\globalMaxSpanHighlight{globalMaxSpan}}.

\noindent\textbf{Description of the properties of span in PD Ontology:} Recall that span is implemented as either the list or the set model, and the PD ontology is expressed using either OWL-DL or OWL-RL. These combinations lead to three versions of the ontology: OWL-DL-list, OWL-DL-set, and OWL-RL-list. The vocabulary used to abstract the span in the list and the set models is already discussed in Section~\ref{sec:pdo}. We now provide an example of the description of a typical property called \textit{non-zero span property} in these three versions.

Recall that OWL-DL ontology can be used in conjunction with SWRL rules. The description of the property in OWL-DL-list version is created using the rules \ref{R1}, \ref{R2} and axioms \ref{A1}, \ref{A2} in Table~\ref{tbl:dlaxioms_span}. The rule \ref{R1} declares that the RDF list \textit{?l1} that has the first cell as zero as an instance of the concept \textit{ListWithZeroElement}. The rule \ref{R2} declares that if the sub-list \textit{?l2} of the RDF list \textit{?l1} belongs to the \textit{ListWithZeroElement} concept, then \textit{?l1} also belongs to the same concept. In these rules, the terminology used in connection with \textit{rdf:List} follows from RDF standard~\cite{rdf}. The axioms \ref{A1} and \ref{A2} gives the definitions of the concepts \textit{ListWithAllNon-ZeroElements} and \textit{SpanWithAllNon-ZeroElements} expressed in OWL-DL. The Span is also a \textit{rdf:List} that is not included in any other list (analogous to the head of a singly linked list). The definition of the non-zero span property in OWL-DL-set version is given in \ref{A3}. The meaning of the axiom is as follows. The \textit{SpanWithAllNon-ZeroElements} concept is defined as the span that has no cell with value zero. In general, the definitions of the properties of span expressed in OWL-DL-set version are concise than \rn{those in} OWL-DL-list version.

%zzz:cite RDF and OWL-RL in the above para
%zzz:Recall that
Recall that OWL-RL is intentionally designed to be less expressive than OWL-DL to preserve the tractability of reasoning. It does not allow SWRL rules and many DL operators such as concept-equivalence. It adopts a peculiar asymmetric syntax in the usage of quantifiers in the class subsumption axioms. The existential quantifier is allowed only on the left-hand side and the universal quantifier only on the right-hand side. In the OWL-RL-list version of the PD ontology, we declared the \textit{ListWithZeroElement} concept as a primitive concept. Even though there is loss of precision, the advantages can be seen in the time taken for reasoning (discussed in Section~\ref{sec:experiment_reasoning}). The non-zero span property is defined by making slight modification to \ref{A1} and \ref{A2} axioms. As the equivalence operator is disallowed, it is replaced with the subsumption operator. The descriptions of the other properties of span are similar to the non-zero span property defined above, \rn{and we omit those for brevity.}

\begin{table*}[t]
\caption{Description Logic axioms and SWRL rules for the non-zero span property in the PD ontologies}
\label{tbl:dlaxioms_span}
\begin{tabular}{ll}
(\text{\newtag{R1}{R1}})   & \textit{rdf:List(?l1) $\wedge$ rdf:first(?l1,?a) $\wedge$ swrlb:equals(?a,0) $\rightarrow$ ListWithZeroElement(?l1)} \\        

(\text{\newtag{R2}{R2}})   & \textit{rdf:List(?l1) $\wedge$ rdf:rest(?l1,?l2) $\wedge$ ListWithZeroElement(?l2) $\rightarrow$ ListWithZeroElement(?l1)} \\        

(\text{\newtag{A1}{A1}}) & \textit{ListWithAllNon-ZeroElements} $\equiv$ \textit{rdf:List} $\sqcap$ $\neg$ \textit{ListWithZeroElement}\\

(\text{\newtag{A2}{A2}}) & \textit{SpanWithAllNon-ZeroElements} $\equiv$ \textit{Span} $\sqcap$ \textit{ListWithAllNon-ZeroElements}\\

(\text{\newtag{A3}{A3}}) & \textit{SpanWithAllNon-ZeroElements} $\equiv$ \textit{Span} $\sqcap$ $\neg$ $\exists$\textit{hasSpanCell} ($\exists$\textit{hasValue} \textit{xsd:long}[$=0$])

\end{tabular}
\end{table*}
\section{Experiments}
\label{sec:experiments}
The BOLD system is implemented in Java and it enabled us to use the existing SPARQL engine~\cite{jena} and Pellet~\cite{pellet} reasoner. We have used SIR~\cite{sir}, a popular bug benchmark repository for experiments. The LOC of the benchmarks \rn{are presented} in Table~\ref{tbl:queryTime}. Each benchmark contains the correct version of the program and several faulty versions seeded with bugs. \rn{We used} the BOLD system to debug faulty versions and \rn{to empirically assess} the performance of different modules. The objective of the assessments is to measure the performance of the main features of BOLD -- querying and reasoning. \rn{We present} the querying-related experiments in Section~\ref{sec:experiment_querying} and the reasoning-related experiments in Section~\ref{sec:experiment_reasoning}. For \rn{faithful} comparison, we have debugged the same faulty versions using \texttt{gdb}.

%We debugged the same version using \texttt{gdb} for comparison.

Debugging is an art, which depends upon the expertise of the user in the profession and the \rn{user's knowledge of} the program. The productivity assessment of debugging involves subjective factors such as guessing the code location that might have caused the bug. For the sake of \rn{automation as well as computing} precision, we restricted our experiments to objective and reproducible parameters. To debug a given faulty program, we have created a sequence of the typical actions a user is likely to perform. We have implemented the same action-sequence in both BOLD and \texttt{gdb}. The action-sequence along with the properties of spans verified for different bugs in this section are explained in the supplementary material. %\dileeptodo{RN to revisit this.}

All the experiments have been performed on a system with Intel Sandy Bridge processor, 10GB RAM, and Ubuntu 16.04 operating system. The reported times are obtained by calculating the \rn{arithmetic} average of five \rn{runs} with the same input. %the specific feature used for comparison.
%\dileeptodo{see if we can move to an i7. By the time your thesis is reviewed, i5 may be considered old. Alternatively, we can mention Intel Xeon processor or whatever it is, but not mention i5.}

\subsection{Performance of Querying}
\label{sec:experiment_querying}
%\dileeptodo{let's also mention about the statistics / information about various queries used in the experiments.}
The querying-related experiments compare the times taken to query the \rn{same} trace information in BOLD \rn{and} in \texttt{gdb}. \rn{For faithful} comparison, we have avoided \rn{any} manual intervention \rn{and have used fully} automated versions of the two systems. A series of commands \rn{is} provided to both the systems at startup. 

%\rn{On the other hand,} in \texttt{gdb}, the script containing the commands that sets the breakpoints at the required \rn{places} is also provided at the startup. %a parameter. %\dileeptodo{Let's make it concrete by showing an example of the commands and the corresponding script.}

\rn{Since our instrumentation and querying using SPARQL may add overheads, it is imperative to check how BOLD's trace generation, information gathering, and querying perform against the well-optimized \texttt{gdb}. We observe that BOLD adds only a minimal overhead, as discussed below.}
Time taken for querying in both the systems is presented in Table~\ref{tbl:queryTime}. The \texttt{gdb-Debug} column indicates the time consumed by \texttt{gdb}. The \texttt{BOLD-Query} column indicates the time taken to generate the trace and query the trace for the relevant information. Since \texttt{gdb} actually executes the program, \rn{to be fair}, we have included the time taken to generate trace in \texttt{BOLD-Query} time \rn{(trace generation also requires program execution)}. %This provides a fair comparison between both the systems. 
From the \texttt{gdb-Debug} and \texttt{BOLD-Query} columns it can be seen that both the systems \rn{exhibit comparable} performance. In all the benchmarks except \textit{printtokens2}, one of the systems performs better than the other only by a \rn{small} margin (order of milliseconds). Considerably larger time in the \textit{printtokens2} benchmark is due to the use of an inter-span property (\textit{isEqualSpanOf}). \rn{Computation of this property builds upon} the results of two SPARQL queries. \rn{Hence, \texttt{BOLD-Query} takes longer. However, we note that the extra time helps} avoid manual labor as in \texttt{gdb}. In \texttt{gdb}, it is a manual task to compare the information at two distinct instances of the statements \rn{, which is not accounted for in \texttt{gdb-Debug} time}.

%In \texttt{gdb}, the equivalent information isn't presented in an formatted manner that .

The last column \rn{presents} the time taken for abstraction in BOLD. This time includes the \texttt{BOLD-Query} time because the abstraction is performed on the results of the queries. It can be seen that the major part of this time involves querying. For all the benchmarks except \textit{flex}, the time taken for abstraction excluding querying is almost constant. The \textit{flex} benchmark does not require abstractions because the debug-operation is to find the program path taken in a function during the execution. The path can be retrieved by integrating the source-code triples (for identifying the branch statements) and execution-trace triples. Finally, the operation is formulated as an integrated query (explained in Section~\ref{sec:querying}). %\dileeptodo{needs some clarity.}

%The querying feature is available in \texttt{gdb}. So we compared the performance of BOLD and \texttt{gdb} using SIR benchmarks

\subsection{Performance of reasoning}
\label{sec:experiment_reasoning}
%The reasoning-related experiments compare the performance of BOLD by varying different sublanguages of OWL and the structures that have been used to model the span. We experimented with two sublanguages of OWL and the two models of the span. Recall that the expressive OWL-DL language requires exponential time and the the restricted OWL-RL language requires only polynomial time for reasoning. Also recall from Section~\ref{sec:paontology} that the list and set models are used to implement the spans.

The reasoning-related experiments compare the performance of BOLD in the three versions of the PD ontology described in Section~\ref{sec:reasoningImplementation}. \rn{Recall that such a reasoning is not possible with the existing debuggers.} The time taken for reasoning the properties of the spans constructed to debug the SIR benchmarks is given in Table~\ref{tbl:reasoningTime}. These spans are constructed over the abstractions of the benchmarks reported in Table~\ref{tbl:queryTime}. The \texttt{Span size} column indicates the number of elements in the span. The \texttt{OWL-DL-list version} and the \texttt{OWL-DL-set version} columns indicate the time taken for reasoning when the span is modelled as list and set respectively, using the OWL-DL language. The last column \texttt{OWL-RL-list version} indicates the reasoning time when the span is modelled as list, using the OWL-RL language. When the span size is small ($\le2$), the set version consumes slightly more time than the OWL-DL-list version. When the span size is large (we call these as big spans), the set version performs considerably faster than the OWL-DL-list version. %The possible reason for this is explained below. 
\rn{Our analysis for this performance difference is presented below.}

The properties verified using the big spans in our experiments are all either universal checks (all elements in the span are same / all are non-zero) or comparison checks (increasing span). In the set version, a negative evidence is sufficient to prove these checks whereas in the list version, the reasoner has to traverse through all the elements because the RDF lists are essentially recursive lists. The axioms and rules explained in Table~\ref{tbl:dlaxioms_span} and Section~\ref{sec:reasoningImplementation} for the non-zero span property provide sufficient justification. \rn{The last version,} OWL-RL-list, clearly outperforms both the OWL-DL versions because of the use of OWL-RL ontology. The complexity of Pellet~\cite{pellet} reasoner used for reasoning depends on the ontology. When used with OWL-RL ontology, it shows tractable performance. 

%The experiments have used a OWL-DL reasoner, called Pellet~\cite{pellet}. When the reasoner is used with OWL-RL ontology, it showed tractable performance.

%because the reasoner is tractable. We have used Pellet~\cite{pellet} reasoner

%\dileeptodo{Write a couple of more lines about the technical details of the reasoner.}

%For example, consider a span $S$ [zzz:Dileep to write]. But the drawback is that the OWL-RL language isn't as expressive as the OWL-DL language. [zzz: Is it already explained?].

It can be seen in Table~\ref{tbl:reasoningTime} that as the span size increases, the performance of BOLD using different versions changes \rn{differently}. So to test the \rn{scalability} of each version, we increased the size of span on a particular benchmark and verified (an arbitrary) property:  \textit{all elements are non-zero}. We enforced a time-out of 2 minutes. The results of the experiment are shown in Table~\ref{tbl:scalingTime}. The \rn{two} versions with OWL-DL language timed-out for span sizes above 100. The version with OWL-RL language clearly shows effective response time even for bigger spans.

Debugging is an \rn{interactive} activity, and the response time of the system is crucial for practical usage. \rn{Our experiments reveal that OWL-RL language provides good performance at scale. OWL-DL language can be used if the span sizes are small.}

%The conclusion is OWL-RL language is recommended in general and OWL-DL language can be used if the size of the spans isn't going to be big.

\begin{table}[t]
\caption{Time taken for querying in \texttt{gdb} and BOLD on SIR benchmarks} %\dileeptodo{Does it make sense to reduce querying time from abstraction column?}}
\label{tbl:queryTime}
\begin{tabular}{|l|r|c|c|c|}
%\begin{tabular}{|l|l|l|l|l|}
\hline
%\textbf{Benchmark} & \textbf{LOC} & \textbf{\begin{tabular}[c]{@{}l@{}}gdb-\\ Debugging\end{tabular}} & \textbf{\begin{tabular}[c]{@{}l@{}}BOLD-\\ Querying\end{tabular}} & \textbf{\begin{tabular}[c]{@{}l@{}}BOLD-\\ Abstraction\end{tabular}} \\ \hline

\multirow{2}{*}{\textbf{Benchmark}} &
\multirow{2}{*}{\textbf{LOC}} &%\multirow{2}{*}{\textbf{\begin{tabular}[c]{@{}c@{}}Span \\ \\ Size\end{tabular}}} & 
\multicolumn{3}{c|}{\textbf{Execution time (in msec)}}                            \\ %\cline{3-5} 
&  &\begin{tabular}[c]{@{}c@{}}\textbf{\texttt{gdb-}}\\ \textbf{Debug}\end{tabular} & \begin{tabular}[c]{@{}c@{}}\textbf{BOLD-}\\ \textbf{Query}\end{tabular} & \begin{tabular}[c]{@{}c@{}}\textbf{BOLD-}\\ \textbf{Abstraction}\end{tabular} \\ \hline

flex&14034&210&290&NA \\ %\hline
grep&10929&205&193&220 \\ %\hline
printtokens&563&206&236&280 \\ %\hline
printtokens2&510&235&637&733 \\ %\hline

schdeule&412&210&187&218 \\ %\hline

sed&8059&210&241&274 \\ %\hline
space&9126&232&203&229 \\ %\hline

totinfo&406&231&202&237 \\ \hline
\end{tabular}
\end{table}

\begin{table}[t]
\caption{Time taken for reasoning in BOLD by using different sub-languages of OWL}
\label{tbl:reasoningTime}
\begin{tabular}{|l|r|r|r|c|}
\hline
%\textbf{Benchmark} & \textbf{\begin{tabular}[c]{@{}l@{}}Span\\ Size\end{tabular}} & \textbf{\begin{tabular}[c]{@{}l@{}}OWL-DL\\ List Model\end{tabular}} & \textbf{\begin{tabular}[c]{@{}l@{}}OWL-DL\\ Set Model\end{tabular}} & \textbf{\begin{tabular}[c]{@{}l@{}}OWL-RL\\ List Model\end{tabular}} \\ \hline
\multirow{2}{*}{\textbf{Benchmark}} & \multirow{2}{*}{\textbf{\begin{tabular}[c]{@{}c@{}}Span \\ size\end{tabular}}} & \multicolumn{3}{c|}{\textbf{Execution time (in msec)}}                                                                       \\ %\cline{3-5} 
& & 
\textbf{\begin{tabular}[c]{@{}c@{}}OWL-DL-\\ list version\end{tabular}} & \textbf{\begin{tabular}[c]{@{}c@{}}OWL-DL-\\ set version\end{tabular}} & \textbf{\begin{tabular}[c]{@{}c@{}}OWL-RL-\\ list version\end{tabular}} \\ \hline

grep&1&354&452&382 \\ %\hline

printtokens&1&390&475&388 \\ %\hline
printtokens2&12&10091&3300&631 \\ %\hline

schedule&7&1411&720&548 \\ %\hline

sed&2&460&500&402 \\ %\hline
space&2&425&520&397 \\ %\hline

totinfo&6&1250&809&507 \\ \hline
\end{tabular}
\end{table}

\begin{table}[t]
\caption{Scalability of reasoning in BOLD by using different sub-languages of OWL (Timeout = 120 seconds)}
\label{tbl:scalingTime}
\begin{tabular}{|r|r|r|r|}
\hline
 %\textbf{\begin{tabular}[c]{@{}l@{}}Span\\ Size\end{tabular}} & \textbf{\begin{tabular}[c]{@{}l@{}}OWL-DL\\ List Model\end{tabular}} & \textbf{\begin{tabular}[c]{@{}l@{}}OWL-DL\\ Set Model\end{tabular}} & \textbf{\begin{tabular}[c]{@{}l@{}}OWL-RL\\ List Model\end{tabular}} \\ \hline
 %\multirow{2}{*}{\textbf{Benchmark}} &
 \multirow{2}{*}{\textbf{\begin{tabular}[c]{@{}c@{}}Span \\ size\end{tabular}}} & \multicolumn{3}{c|}{\textbf{Execution time (in sec)}}                                                                       \\ %\cline{2-4} 
 &
 \textbf{OWL-DL-list} & \textbf{OWL-DL-set} & \textbf{OWL-RL-list} \\ \hline
10&2.5&1.1&0.22 \\ %\hline
30&28.1&2.1&0.23 \\ %\hline
50&115.8&19.1&0.24 \\ %\hline
100&Timeout&32.1&0.27 \\ %\hline
200&Timeout&Timeout&0.32 \\ %\hline
1000&Timeout&Timeout&0.78 \\ %\hline
5000&Timeout&Timeout&16.23 \\ \hline
\end{tabular}
\end{table}

%\newpage

% total 58 lines per page. first category- 16 lines. second-18 lines The third category took 13 lines. avg (19 lines)

\section{Related work}
\label{sec:relatedWork}
We discuss related work in three categories. First, we discuss the existing approaches for instrumentation. The source code preprocessors (\rn{such as} \texttt{gcc}, Rose~\cite{rose_instrumentation}, OPARI~\cite{opari_instrumentation}) typically provide directives to control instrumentation. They focus on a specific issue or uses hard-coded commands in the program. This affects the readability and maintenance of the code. Many tools~\cite{custom1_instrumentation,custom2_instrumentation,custom3_instrumentation} provide custom methods to control instrumentation. These tools lack a standard terminology to define the instrumentation specification. Aspect-oriented programming~\cite{aop_instrumentation} generalizes functional decomposition and separates aspect code from the \rn{original} code. This paradigm is successfully used for instrumentation by \cite{aop_cpp_instrumentation}. But, the drawback is that it is tied to the programming language and does not provide standardization. There are languages (\rn{such as} PTQL~\cite{ptql_instrumentation}, PQL~\cite{pql_instrumentation}, UFO~\cite{ufo_instrumentation}) designed to instrument queries into the source code. In principle, BOLD offers a \rn{comprehensive} solution than these languages. \rn{Further,} unlike these languages, \rn{in BOLD,} programs need not be recompiled for each new query.

Second, we discuss the existing log-based debuggers and the tools that support high-level abstractions over the trace. The ODB system~\cite{odb} designed for Java programs uses \textit{ad hoc} format to store the trace. The Tralfamadore~\cite{tralfamadore} system is based on the ideas of streaming databases. The TOD~\cite{pothier} system uses distributed databases to support storage and query processing. These systems provide an interface to navigate through sub-parts of the execution trace. The replay-based debuggers log information at specific places called \textit{checkpoints}. Some examples include UndoDB~\cite{undodb} and URDB~\cite{urdb} for user-level programs. A survey on replay-based debugging can be found in Engblom~\cite{engblom}. \rn{On the other hand,} some tools provide high-level abstractions to simplify debugging. The Expositor tool~\cite{expositor} abstracts the execution trace as a list and the users can query the trace using list processing API. The MzTake~\cite{mztake}, Dalek~\cite{dalek}, EBBA~\cite{ebba} systems adopt a similar approach. The Coca~\cite{coca} system allows users to write Prolog predicates over program states. Overall, these debuggers lack a standard semantic model to store and analyse the execution trace in a standard manner.
%odb,tralfamadore,pothier,engblomSystem,undodb
%expositor,mztake,dalek,ebba,coca

Third, we discuss the application of ontologies in different problems related to programming. PATO~\cite{ecoop} and CodeOntology~\cite{codeontology} frameworks proposed ontologies for C and Java languages respectively. The PATO framework~\cite{ecoop} converts C programs into triples and performs static analysis using those. The SmartAPI~\cite{smartapi} approach uses ontologies for automatic code generation. They expand libraries with domain ontology which assist easy location of code for a given goal. Ontologies are used to publish meta-information of software~\cite{swo}. This caters to the needs of a broad range of users. Ontologies are used for code maintenance~\cite{lassie} and teaching programming languages~\cite{ontologyteaching1,ontologyteaching2}. The applications of ontologies in software engineering are discussed by \cite{sereference}.

%\rn{In summary, BOLD differs from the existing works by providing a standard and extensible, ontology-based semantic model for debugging.}

\rn{In summary, BOLD framework differs from the existing works by providing a standard and extensible, ontology-based semantic model for debugging.}
\section{Conclusion and future work}
\label{sec:conclusion}
We presented a unified framework called BOLD for debugging. The framework integrates different tasks in debugging such as instrumentation, representation, and analysis of execution trace in a standard manner. The standardization is accomplished using an OWL ontology called PD ontology. The ontology provides different properties of spans (sub-sequences of trace list) to simplify debugging. We have used the \rn{popular} SIR benchmarks to demonstrate the effectiveness of the framework. The experiments indicate that we are able to diagnose the causes of the bugs in SIR benchmarks. The time taken is also practical for an interactive session. %manual effort involved  BOLD has diagnosed the causes of SIR bugs in reasonable time. %Additionally, the the time taken to query the trace is comparable to \texttt{gdb}. 
%\dileeptodo{mention the last line in more detail. mention about reasoning.}
%, using the properties of spans provided in the PD ontology,

A very interesting direction for future work is to extend the BOLD framework to create and reason about program assertions. The current approaches for assertions are restricted to a statement. Ontologies provide a general way to create assertions about the whole programs. These assertions can be used to automatically identify and generate explanations for the bugs.
%\rn{This is a very interesting direction.}
\bibliographystyle{unsrt}
\bibliography{main}

\end{document}